\documentclass[twocolumn,english,showpacs,floatfix]{revtex4}
\usepackage[dvips]{graphicx}
\usepackage{longtable}
\usepackage{amsmath,amssymb}
\usepackage{dcolumn}

\usepackage{babel}
\usepackage[latin1]{inputenc}

\begin{document}


\title{Quantization of  massive scalar fields
over static black string backgrounds}
\author{Owen Pavel Fern\'{a}ndez Piedra}
\email{opavel@ucf.edu.cu }
\affiliation{Departamento de F\'{i}sica y Qu\'{i}mica, Universidad de Cienfuegos, Carretera a Rodas, Cuatro Caminos, s/n. Cienfuegos, Cuba}

\author{Alejandro Cabo Montes de Oca}
\email{cabo@icmf.inf.cu}
\affiliation{Grupo de F\'{i}sica Te\'{o}rica, ICIMAF, Calle E  No. 309, esq. a 15 Vedado, C. Habana,Cuba.}

\date{\today}

\begin{abstract}

 \noindent The renormalized mean value of the corresponding components of the
Energy-Momentum tensor for massive scalar fields coupled to an arbitrary gravitational field configuration having cylindrical symmetry  are analytically evaluated using the Schwinger-DeWitt
approximation, up to second order in the inverse mass value. The general results are employed to explicitly derive compact analytical expressions for the Energy-Momentum tensor in the particular
background of the Black-String space-time. In the case of the Black String considered in this work, we proof that a violation of the weak energy condition occur at the horizon of the space-time for
values of the coupling constant, that include as particular cases the most interesting of minimal and conformal coupling.
\end{abstract}


\maketitle

Semiclassical gravity considers the quantum dynamics of fields in a
gravitational background, which  at this level of description is
considered as a classical external field. That is, all fields are
considered as quantum ones, with the only exception of the external
gravitational field, that remains satisfying the classical Einstein
field equations, associated with sources given by the vacuum
expectation values of the stress energy tensor of the matter fields
\cite{birrel-fulling}. In that situation, it is necessary to have
adequate mathematical methods  to obtain explicit analytical
expressions for the renormalized stress tensor \(\langle
T_{\mu}^{\nu}\rangle_{ren}\), the quantity that enters as a source
in the semiclassical Einstein equations \cite{frolov-bv,
DeWitt-gilkey, avramidi,exactresults,numeric, matyjasek1,matyjasek}.
This stress tensor and the expectation value
\(\langle\varphi^{2}\rangle_{ren}\) of a quantum field \(\varphi\)
are the main objects to calculate from quantum field theory in
curved spacetime.

Having the components for \(\langle T_{\mu}^{\nu}\rangle_{ren}\), the backreaction of the quantized fields in the space-time geometry of black holes can in principle be determined, unless the
(unknown) effects of quantum gravity become important. There are several approximate methods to build the effective action, starting from which the energy momentum tensor can be calculated by
functional differentiation with respect to the metric \cite{exactresults}. One of the developed techniques, the so called Schwinger-De Witt expansion, is based on a series development expansion of
the effective action in inverse powers of the field mass. It is well-known that this method can be used to investigate effects like the vacuum polarization of massive fields in curved backgrounds,
whenever the Compton's wavelength of the field is less than the characteristic radius of curvature \cite{frolov-bv, DeWitt-gilkey, avramidi, matyjasek}. Also numerical computations of \(\langle
T_{\mu}^{\nu}\rangle_{ren}\) and \(\langle\varphi^{2}\rangle_{ren}\) have been performed by a number of authors \cite{numeric}.

In General Relativity there exists a four parameter family of black hole solutions called the generalized Kerr-Newmann family. The solutions belonging to this family are characterized by the four
parameters:  mass \(M\), angular momentum \(J\), charge \(Q\) and the Cosmological Constant \(\Lambda\) \cite{lemos}. These are axis-symmetric solutions that show different asymptotic behavior
depending on the sign of the cosmological constant. There are two important cases of axial symmetry. One is the spherical symmetry that have been studied in great detail since the birth of General
Relativity. The other one is the cylindrical symmetry. As it has been shown by Lemos in Ref. \cite{lemos}, in the case of negative cosmological constant, there exists a Black hole solution showing
cylindrical symmetry: the  so called Black String. Charged rotating black string solutions has many similarities with the Kerr-Newman black hole, apart from space-time being asymptotically anti-de
Sitter in the radial direction (and not asymptotically flat). The existence of black strings suggests that they could be the final state of  the collapse of matter having cylindrical symmetry.

In a cylindrical coordinate system \((x^{0},x^{1},x^{2},x^{3})=(t,\rho,\varphi,z)\) with \(-\infty<t<\infty\), \(0\leq \rho<\infty\), \(-\infty<z<\infty\), \(0\leq \varphi<\infty\),the metric of the
static black string showing both charge and angular momentum equal to zero is (see Ref. \cite{lemos}):
\begin{equation}\label{}
	ds^{2}=(\frac{4M}{\alpha\rho}-\alpha^{2}\rho^{2})dt^{2}+
	\frac{d\rho^{2}}{(\alpha^{2}\rho^{2}-\frac{4M}{\alpha\rho})}+
	\rho^{2}d\varphi^{2}+(\alpha\rho)^{2}dz^{2}  \label{ds2simp}
\end{equation}
with the constant \(\alpha\) defined as \(\alpha^{2}=-\frac{1}{3}\Lambda,\) where \(\Lambda\) is a negative  Cosmological Constant. The above metric has an event horizon located at
\(\rho_{H}=\frac{\sqrt[3]{4M}}{\alpha}\). The apparent singular behavior at this horizon is a coordinate effect and not a true one. The only true singularity is a polynomial one at the origin.

Previous works on the problems of determining \(\langle\varphi^{2}\rangle_{ren}\) and the renormalized stress tensor components for conformally coupled massless scalar fields in black String
backgrounds and the calculation of gravitational backreaction effects of the quantum field were performed by DeBenedictis in \cite{debenedictis}.

In the following we use for the Riemann tensor, its contractions, and the covariant derivatives the sign conventions of Misner, Thorne and Wheeler \cite{misner}. Our units are such that
\(\hbar=c=G=1\).

In this paper we derive the exact expressions for the components of the renormalized stress tensor for  massive scalar field in the black string background. We first consider the derivation of the
Schwinger-De Witt approximation for the renormalized Lagrangian and Stress-Energy tensor of massive scalar field subject to an arbitrary background space-time. Consider a single massive scalar field
\(\phi(x)\) with mass $m$ interacting with gravity with non minimal coupling constant \(\xi\) in four dimensions. The action is:

\begin{equation}\label{}
	S=-\frac{1}{2}\int d^{4}x\sqrt{-g}\left(\nabla^{\mu}\phi\nabla_{\mu}\phi+(m^{2}+\xi R)\phi^{2}\right)
\end{equation}
Using DeWitt's effective action approach and applying Schwinger's regularization prescription \cite{frolov-bv, avramidi}  one gets the renormalized effective action for the quantized scalar field
satisfying equation (1) as
\begin{eqnarray}
W_{ren}&=&\int  d^{4}x \sqrt{-g}\,\mathfrak{L}_{ren}
\nonumber \\
&=&{1\over 2(4\pi)^{2}\,} \int  d^{4}x \sqrt{-g}\,\sum_{k=3}^{\infty}{\,str \,a_{k}(x,x)\over k(k-1)(k-2)m^{2(k-2)}}
\nonumber \\
\end{eqnarray}
In the above expression \(str F\) denotes the functional supertrace of \(F\) \cite{avramidi}. The coefficients \([a_{k}]= \,a_{k}(x,x')\), whose coincidence limit appears under the supertrace
operation in (7) are the Hadamard-Minakshisundaram-DeWitt-Seeley coefficients (HMDS). As usual, the first three coefficients of the DeWitt-Schwinger expansion, $a_{0},\,a_{1},\,{\rm and}\,a_{2}, $
contribute to the divergent part of the action and can be absorbed in the classical gravitational action by renormalization of the bare gravitational and cosmological constants. Various authors have
calculated some of the HDSM coefficients in exact form up to $n \geq 4$ \cite{DeWitt-gilkey,avramidi}. In this report we use the first non-vanishing term of the renormalized effective action above
as our approximate effective action. As was proved in \cite{frolov-bv} the differential operators related with the equation of motion for fields of higher spin can be manipulated in a form that
results in analogous effective actions differing only by numerical coefficients, and one can generalize the presented results to fields of other spins. However, it is important to stress that the
method is restricted  to cases of massive fields in which we avoid the presence of strong or rapidly varying gravitational fields. Restricting ourselves here to the terms proportional to $m^{-2},$
\cite{avramidi,matyjasek}, we obtain an expression for the renormalized effective lagrangian involving only geometrical terms associated with the background space-time, suitable for the calculation
of the stress tensor by functional derivation with respect to the metric. We refer the reader to the complete formulas in \cite{owen}. The renormalized Stress-Energy tensor can be written in a
general form as \(\langle T_{\mu}^{\ \ \nu}\rangle_{ren}=C_{\mu}^{\ \ \nu}+D_{\mu}^{\ \ \nu},\) where the $C_{\mu}^{\ \ \nu}$ and $D_{\mu}^{\ \ \nu}$ tensors take the somewhat cumbersome forms:
\begin{widetext}
\begin{eqnarray*}
C_{\mu}^{\ \ \nu} &=&\frac{1}{96\pi^{2} m^{2}}\left[\frac{1}{7560}\Theta ( \nabla_{\mu}R\nabla^{\nu}R\,+\,\nabla^{\nu}\nabla_{\mu}(\Box R)\,+\,\nabla_{\mu}\nabla^{\nu}(\Box R)\,-\,2 \Box^{2} R
\delta_{\mu}^{\ \ \nu}\,-\ {1\over 2} \delta_{\mu}^{\ \ \nu}\nabla_{\gamma}R\nabla^{\gamma}R\,-\,2 \Box R \nabla^{\nu}\nabla_{\mu}R ) \right.
\nonumber \\
&&\left. \,+\,{1\over 140}\left[\nabla_{\mu}R_{\gamma \lambda}\nabla^{\nu}R^{\gamma \lambda}\,-\, \nabla^{\nu}R_{\gamma \lambda}\nabla^{\lambda}R^{\ \ \gamma}_{\mu}\,-\, \nabla_{\mu}R_{\gamma
\lambda}\nabla^{\lambda}R^{\gamma \nu}\,+ \nabla^{\gamma}R_{\gamma \lambda}\nabla^{\nu}R^{\ \ \lambda}_{\mu}\,+\,\nabla^{\gamma}R_{\gamma \lambda}\nabla_{\mu}R^{\lambda \nu}\,+\,
\nabla^{\gamma}\nabla^{\nu}(\Box R_{\gamma \mu})\right.\right.
\nonumber \\
&&\left.\left.\,-\,\Box^{2} R_{\mu}^{\ \ \nu} \,+\nabla^{\gamma}\nabla_{\mu}(\Box R_{\gamma}^{\ \nu}) \,-\, {1\over 2} \nabla_{\varrho}R_{\gamma \lambda}\nabla^{\varrho} R^{\gamma
\lambda}\delta_{\mu}^{\ \nu}\,- \nabla^{\gamma}\nabla^{\lambda}(\Box R_{\gamma \lambda})\delta_{\mu}^{\ \nu}\,+\,\nabla_{\lambda}\nabla^{\nu}R_{\gamma \mu}R^{\gamma \lambda} \,+\,
\nabla_{\lambda}\nabla_{\mu}R_{\gamma}^{\ \nu} R^{\gamma \lambda}\right.\right.
\nonumber \\
&&\left.\left.\,-\ \nabla^{\gamma}\nabla^{\nu}R_{\gamma \lambda}R^{\ \ \gamma}_{\mu}\,-\,
 \left(\nabla_{\lambda} \nabla_{\sigma} R_{\ \gamma}^{\lambda \ \ \sigma \nu}\,+\,{1\over 2} \nabla^{\nu} \nabla_{\gamma}R \,-\, R^{\lambda \sigma}R_{\gamma \ \lambda \sigma}^{\ \nu}\right)R^{\ \ \gamma}_{\mu}
\,+\, R_{\gamma\lambda}R^{\gamma\lambda}R^{\ \ \gamma}_{\mu} \,-\, \Box R_{\gamma \mu}R^{\gamma \nu} \right.\right.
\nonumber \\
&&\left.\left.\,- \nabla^{\lambda}\nabla_{\mu}R_{\gamma \lambda}R^{\gamma \nu}\right]\,-\,{24\over 1890}\left[ \nabla^{\nu}R_{\gamma \lambda}\nabla^{\lambda} R^{ \ \ \gamma}_{\mu} \,+\,
\nabla^{\gamma} R_{\gamma \lambda}\nabla_{\mu}R^{\lambda \nu}\,-\, \nabla^{\gamma} R_{\gamma \lambda} \nabla^{\varrho} R_{\varrho}^{ \ \lambda } \delta_{\mu}^{\ \nu}\,-\, \nabla_{\varrho}R_{\gamma
\lambda} \nabla^{\lambda} R^{ \gamma \varrho} \delta_{\mu}^{\ \nu}\right.\right.
\nonumber \\
&&\left.\left.\,+\,\nabla_{\lambda} \nabla^{\nu}R_{\gamma \mu}R^{\gamma \lambda}\,+\, \nabla_{\lambda} \nabla_{\mu}R_{\gamma}^{\ \nu} R^{\gamma \lambda}\,-\, \nabla^{\lambda}
\nabla_{\varrho}R_{\gamma \lambda}R^{\gamma \varrho} \delta_{\mu}^{ \ \nu}\,+\, \nabla^{\lambda} \nabla^{\nu}R_{\gamma \lambda} R^{ \ \ \gamma}_{\mu} \,-\,\left(\nabla_{\lambda} \nabla_{\sigma} R_{\
\gamma}^{\lambda \ \ \sigma \nu}\,+\,{1\over 3} \nabla^{\nu} \nabla_{\gamma}R\right.\right.\right.
\nonumber \\
&&\left.\left.\left. \,-\,{2\over 3} R^{\lambda \sigma}R_{\gamma \ \lambda \sigma}^{\ \nu}\,+\,{2\over 3} R_{\gamma\lambda}R^{\gamma\lambda}\right) R^{\ \ \gamma}_{\mu} \,+\,\nabla^{\lambda}
\nabla_{\mu}R_{\gamma \lambda} R^{\gamma \nu}\,-\, \Box R_{\gamma \mu} R^{\gamma \nu} \,-\, \nabla_{\varrho} \nabla^{\gamma}R_{\gamma \lambda} R^{\lambda \varrho} \delta_{\mu}^{ \ \nu}\,+\,R_{\gamma
\lambda} R_{\varrho}^{\ \gamma} R^{\lambda \varrho}\delta_{\mu}^{\ \nu}\right.\right.
\nonumber \\
&&\left.\left. \,-\, 2 R_{\gamma \lambda} R^{\ \ \gamma}_{\mu} R^{\lambda \nu}\,+\,\nabla_{\mu}R_{\gamma \lambda} \nabla^{\lambda} R^{\gamma \nu} \,-\, 2 \nabla_{\gamma} R_{\gamma \mu}
\nabla^{\lambda} R^{\gamma \nu}\,+\, \nabla^{\gamma}R_{\gamma \lambda} \nabla^{\nu} R^{\ \ \lambda}_{\mu}\right]\,+\,{2\over 315}\left(\nabla^{\gamma}R_{\gamma \mu}\nabla_{\lambda} R_{\lambda}^{\
\nu}\right.\right.
\nonumber \\
&&\left.\left.\,+\,\nabla_{\lambda}R_{\gamma}^{\ \nu}\nabla^{\gamma} R^{\ \ \lambda}_{\mu}\,-\, 2 \nabla^{\gamma}R_{\gamma \lambda} \nabla^{\lambda} R_{\mu}^{\ \ \nu}\,-\,\nabla^{\nu} R_{\gamma
\lambda} \nabla^{\varrho} R_{\varrho \ \ \mu}^{\ \gamma \lambda}\,+\,\nabla_{\varrho}R_{\gamma \lambda} \nabla_{\mu}R^{\gamma \varrho \lambda \nu} \,+\, 2 \nabla_{\varrho}R_{\gamma \lambda}
\nabla^{\sigma}R_{\sigma}^{\ \gamma \lambda \varrho } \delta_{\mu}^{\ \nu} \right.\right.
\nonumber \\
&&\left.\left.\,-\,\nabla_{\lambda} \nabla_{\gamma} R_{\mu}^{\ \ \nu}R^{\gamma \lambda}\,+\, \nabla^{\varrho} \nabla^{\nu}R_{\gamma \lambda \varrho \mu}R^{\gamma \lambda}\,-\,\Box R_{\gamma \mu
\lambda}^{\ \ \ \ \nu} R^{\gamma \lambda}\,+\,\nabla^{\varrho} \nabla_{\mu}R_{\gamma \ \lambda \varrho} ^{\ \nu} R^{\gamma \lambda}\,-\,\nabla^{\lambda} \nabla^{\sigma}R_{\gamma \lambda \varrho
\sigma} R^{\gamma \varrho} \delta_{\mu}^{\ \nu}\right.\right.
\nonumber \\
\end{eqnarray*}
\begin{eqnarray}
&&\left.\left.\,+\, {1\over 2} \nabla^{\gamma} \nabla_{\lambda}R_{\gamma}^{\ \ \nu} R^{\ \ \lambda}_{\mu}\,+\,{1\over 2} \nabla_{\lambda} \nabla^{\gamma}R_{\gamma}^{\ \nu}R^{ \ \ \lambda}_{\mu}
\,+\, {1\over 2} \nabla^{\gamma} \nabla_{\lambda}R_{\gamma \mu} R^{\lambda \nu}\,+\, {1\over 2} \nabla_{\lambda} \nabla^{\gamma}R_{\gamma \mu} R^{\lambda \nu}\,+\, {1\over 2} R_{\gamma \lambda}
R_{\varrho \sigma} R^{\gamma \varrho \lambda \sigma} \delta_{\mu}^{\ \nu} \right.\right.
\nonumber \\
&&\left.\left.\,-\, {3\over 2} R_{\gamma \lambda} R_{\varrho \mu} R^{\gamma \varrho \lambda \nu}\,-\,\nabla^{\gamma} \nabla^{\lambda}R_{\gamma \lambda} R_{\mu}^{\ \ \nu} \,+\, \nabla_{\varrho}
\nabla^{\nu}R_{\gamma \lambda} R^{\gamma \varrho \lambda}_{\ \ \ \mu}\,+\, \nabla_{\varrho} \nabla_{\mu}R_{\gamma \lambda} R^{\gamma \varrho \lambda \nu}\,-\,\nabla_{\sigma} \nabla_{\varrho
}R_{\gamma \lambda} R^{\gamma \sigma \lambda \varrho} \delta_{\mu}^{\ \nu}\right.\right.
\nonumber \\
&&\left.\left.\,-\, \Box R_{\gamma \lambda} R^{\gamma \ \ \lambda \nu}_{\ \mu}\,-\, \nabla_{\mu} R_{\gamma \lambda} \nabla^{\varrho} R_{\varrho}^{\ \gamma \lambda \nu}\,+\,\nabla_{\varrho}R_{\gamma
\lambda} \nabla^{\nu} R^{\ \ \gamma \varrho \lambda }_{\mu}\,-\,2 \nabla_{\varrho} R_{\gamma \lambda} \nabla^{\varrho}R^{\gamma \ \ \lambda \nu}_{\ \mu}\right)\,-\,{51\over 7560}\left(- 2
\nabla^{\varrho} R_{\gamma \lambda \varrho}^{\ \ \ \nu} \nabla^{\sigma} R_{\sigma \mu}^{\ \ \ \gamma \lambda }\right.\right.
\nonumber \\
&&\left.\left.\,+\,\nabla^{\varrho} \nabla_{\sigma} R_{\gamma \lambda \varrho}^{\ \ \ \nu} R^{\gamma \lambda \sigma}_{ \ \ \ \mu} \,-\, \nabla^{\varrho} \nabla_{\sigma} R_{\gamma \lambda \varrho
\mu} R^{\gamma \lambda \sigma \nu}\,-\, \nabla_{\sigma} \nabla^{\gamma} R_{\gamma \ \lambda \varrho}^{\ \nu} R^{\lambda \varrho \sigma}_{ \ \ \ \mu}\,-\,R_{\gamma \lambda \varrho \mu} R_{\sigma
\tau}^{\ \ \varrho \nu} R^{\gamma \lambda \sigma \tau}\,+\, {1\over 6} R_{\gamma \lambda \varrho \sigma} R_{\tau \chi}^{\ \ \gamma \lambda} R^{\varrho \sigma \tau \chi} \delta_{\mu}^{\
\nu}\right.\right.
\nonumber \\
&&\left.\left.\,-\,\nabla_{\sigma} \nabla^{\gamma} R_{\gamma \mu \lambda \varrho }R^{\lambda \varrho \sigma \nu}\,-\, 2\nabla_{\sigma} R_{\gamma \lambda \varrho }^{\ \ \ \nu}
\nabla^{\varrho}R^{\gamma \lambda \sigma}_{\ \ \ \mu}\right)+ {1\over 1260}\left(\,-\,{1\over 2} \Box R_{\gamma \mu \lambda \varrho} R^{\gamma \nu \lambda \varrho}\,+\,2 \nabla_{\lambda}R_{\gamma
\mu} \nabla^{\varrho}R_{\varrho}^{\ \nu \gamma \lambda} \,-\, 2 \nabla^{\gamma}R_{\gamma \lambda} \nabla^{\varrho} R_{\varrho \ \ \mu}^{\ \nu \lambda}\right.\right.
\nonumber \\
&&\left.\left.\,+\,\nabla^{\nu}R_{\gamma \lambda \varrho \sigma} \nabla^{\sigma}R^{\gamma \lambda \varrho}_{\ \ \ \mu}\,-\,\nabla^{\gamma}R_{\gamma \lambda \varrho \sigma} \nabla^{\nu} R^{\lambda \
\ \varrho \sigma}_{\ \mu} \,-\, \nabla_{\sigma}R_{\gamma \lambda \varrho \mu} \nabla^{\sigma} R^{\gamma \lambda \varrho \nu}\,-\,2 \nabla_{\varrho}R_{\gamma \lambda} \nabla^{\lambda} R^{\gamma \ \
\varrho \nu}_{ \ \mu}\,-\, {1\over 2} \nabla^{\gamma} R_{\gamma \lambda \varrho \sigma} \nabla^{\tau} R_{\tau}^{\ \lambda \varrho \sigma} \delta_{\mu}^{\ \nu} \right.\right.
\nonumber \\
&&\left.\left.\,-\,2 \nabla_{\varrho} \nabla^{\gamma} R_{\gamma \ \lambda \mu}^{\ \nu} R^{\lambda \varrho} \,-\, 2 \nabla^{\gamma} \nabla^{\varrho} R_{\gamma \lambda \varrho}^{\ \ \ \nu} R^{ \ \
\lambda}_{\mu}\,+\,2 \nabla_{\varrho} \nabla_{\lambda} R_{\gamma \mu} R^{\gamma \varrho \lambda \nu}\,+\,\nabla_{\sigma} \nabla^{\nu} R_{\gamma \mu \lambda \varrho} R^{\gamma \sigma \lambda \varrho}
\,-\,{1\over 2} \nabla^{\lambda} \nabla_{\tau}R_{\gamma \lambda \varrho \sigma} R^{\gamma \tau \varrho \sigma} \delta_{\mu}^{\ \nu} \right.\right.
\nonumber \\
&&\left.\left.\,+\, R_{\gamma \mu } R_{\lambda \varrho \sigma}^{\ \ \ \nu} R^{\gamma \sigma \lambda \varrho}-2 \nabla_{\lambda}R_{\gamma \mu} \nabla^{\varrho}R_{\varrho}^{\ \gamma \lambda \nu}\,-\,
{1\over 2}\nabla_{\tau} R_{\gamma \lambda \varrho \sigma} \nabla^{\sigma} R^{\gamma \lambda \varrho \tau} \delta_{\mu}^{\ \nu}\,-\,2 \nabla^{\gamma} \nabla_{\varrho} R_{\gamma \lambda} R^{\lambda \
\ \varrho \nu}_{\ \mu} \,+\, {1\over 2} \nabla_{\tau} \nabla^{\gamma} R_{\gamma \lambda \varrho \sigma} R^{\lambda \tau \varrho \sigma} \delta_{\mu}^{\ \nu}\right.\right.
\nonumber \\
&&\left.\left.\,+\, \nabla_{\lambda} \nabla^{\nu} R_{\gamma \lambda \varrho \sigma} R^{\gamma \ \ \varrho \sigma}_{\ \mu}\,-\,{1\over 2} \Box R_{\gamma \ \lambda \varrho }^{\ \nu} R^{\gamma \ \
\lambda \varrho}_{\ \mu}\right)\,-\,{3\over 540}\left(2 \nabla^{\gamma} R_{\gamma \lambda \varrho \mu} \nabla^{\sigma} R_{\sigma}^{\ \varrho \lambda \nu}\,+\, {1\over 3} R_{\gamma \lambda \varrho
\sigma} R_{\tau \ \chi}^{\ \gamma \ \varrho} R^{\lambda \tau \sigma \chi}\delta_{\mu}^{\ \nu}\right.\right.
\nonumber \\
&&\left.\left.\,+\,2 \nabla^{\gamma} R_{\gamma \lambda \varrho \sigma} \nabla^{\sigma} R^{\gamma \nu \varrho}_{ \ \ \ \mu} \,-\,\nabla_{\sigma} \nabla_{\varrho} R_{\gamma \mu \lambda}^{\ \ \ \ \nu}
R^{\gamma \sigma \lambda \varrho}\,-\, \nabla_{\sigma} \nabla_{\varrho} R_{\gamma \ \lambda \ \mu}^{\ \nu } R^{\gamma \sigma \lambda \varrho}\,+\,\nabla_{\sigma} \nabla^{\varrho} R_{\gamma \ \lambda
\varrho }^{\ \nu} R^{\gamma \sigma \lambda }_{\ \ \ \mu}\,+\,\nabla_{\sigma} \nabla^{\varrho} R_{\gamma \mu \lambda \varrho} R^{\gamma \sigma \lambda \nu} \right.\right.
\nonumber \\
&&\left.\left.\,-\, 2 R_{\gamma \lambda \varrho \mu} R_{\sigma \ \tau}^{\ \lambda \ \nu} R^{\gamma \sigma \varrho \tau}\,+\,\nabla_{\sigma} \nabla^{\lambda} R_{\gamma \lambda \varrho }^{\ \ \ \nu}
R^{\gamma \ \  \varrho \sigma}_{ \ \mu}\,-\,\nabla^{\lambda} \nabla^{\sigma} R_{\gamma \lambda \varrho \sigma} R^{\gamma \ \ \varrho \nu}_{\ \mu}\,+\, \nabla^{\lambda} \nabla_{\sigma} R_{\gamma
\lambda \varrho \mu} R^{\gamma \nu \varrho \sigma}\,+\,2 \nabla^{\gamma} R_{\gamma \lambda \varrho \sigma} \nabla^{\sigma} R^{\lambda \ \ \varrho \nu}_{\ \mu}\right.\right.
\nonumber \\
&&\left.\left.\,-\,\nabla^{\lambda} \nabla^{\sigma} R_{\gamma \lambda \varrho \sigma} R^{\gamma \nu \varrho}_{\ \ \ \mu} \,+\,2\nabla_{\sigma} R_{\gamma \lambda \varrho }^{\ \ \ \nu}
\nabla^{\lambda}R^{\gamma \ \ \varrho \sigma}_{\ \mu}\right)\right]. \nonumber \\ \label{emTensor1}
\end{eqnarray}
and:
\begin{eqnarray}
D_{\mu}^{\ \ \nu} &=&\frac{1}{96\pi^{2} m^{2}}\left[{1\over 30}\eta\left( \nabla^{\nu}R \nabla^{\gamma} R_{\gamma \mu}\,+\,\nabla_{\mu}R \nabla^{\gamma}R_{\gamma}^{\ \nu} \,+\,2
\nabla^{\nu}R_{\gamma \lambda}\nabla_{\mu} R^{\gamma \lambda}\,-\, \Box R R_{\mu}^{\ \ \nu}\,+\, \nabla_{\gamma}R \nabla^{\nu}R^{\ \ \gamma}_{\mu} \,+\, \nabla_{\gamma}R \nabla_{\mu} R^{\gamma \nu}
\right.\right.
\nonumber \\
&&\left.\left. \,-\, 2 \nabla_{\gamma}R \nabla^{\gamma} R_{\mu}^{\ \ \nu}\,+\ R \nabla^{\gamma}\nabla^{\nu}R_{\gamma \mu}\,+\,R \nabla^{\gamma}\nabla_{\mu}R_{\gamma}^{\ \nu}-R \nabla_{\lambda}
\nabla_{\gamma} R_{\ \mu}^{\lambda \ \ \gamma \nu}\,-\,{1\over 2}R \nabla^{\nu} \nabla_{\mu}R \,+\,R R^{\lambda \gamma}R_{\mu \lambda \ \gamma}^{\ \ \ \nu}\,-\,R
R_{\mu\lambda}R^{\mu\lambda}\right.\right.
\nonumber \\
&&\left.\left. \,-\ 2 \nabla_{\gamma}R \nabla^{\lambda}R_{\lambda}^{\ \gamma} \delta_{\mu}^{\ \nu}\,-\nabla^{\nu}\nabla_{\mu}R_{\gamma \lambda} R^{\gamma
\lambda}\,+\,\nabla_{\mu}\nabla^{\nu}R_{\gamma \lambda} R^{\gamma \lambda} \,+\, \nabla_{\lambda}\nabla_{\gamma}R R^{\gamma \lambda} \delta_{\mu}^{\ \nu}\,-\ 2 \Box R_{\gamma \lambda} R^{\gamma
\lambda} \delta_{\mu}^{\ \nu} \,+\, {1\over 2} R R_{\gamma \lambda} R^{\gamma \lambda} \delta_{\mu}^{\ \nu}\right.\right.
\nonumber \\
&&\left.\left. \,+\,\nabla^{\nu}\nabla_{\gamma}R R^{\ \ \gamma}_{\mu}\,-\ 2 R R_{\gamma}^{\ \nu} R^{\ \ \gamma}_{\mu} \,+\,\nabla_{\mu}\nabla_{\gamma}R R^{\gamma \nu}\,-\ R_{\gamma \lambda}
R^{\gamma \lambda} R_{\mu}^{\ \ \nu}\,+\,4 \nabla_{\gamma} R \nabla^{\lambda} R_{\lambda \mu}^{\ \ \ \gamma \nu}\,+\,2 \nabla_{\tau}R_{\gamma \lambda \varrho \sigma}\nabla^{\tau} R^{\gamma \lambda
\varrho \sigma } \delta_{\mu}^{\ \nu}\right.\right.
\nonumber \\
&&\left.\left. \,+\,4 \nabla_{\gamma}R \nabla^{\lambda} R_{\lambda \ \ \mu}^{\ \nu \gamma}\,-\, 2 \nabla^{\nu}R_{\gamma \lambda \varrho \sigma}\nabla_{\mu}R^{\gamma \lambda \varrho \sigma }\,+\, 2 R
\nabla^{\gamma}\nabla^{\lambda}R_{\gamma \mu \lambda}^{ \ \ \ \ \nu } \,- \nabla^{\nu} \nabla_{\mu}R_{\gamma \lambda \varrho \sigma}R^{ \gamma \lambda \varrho \sigma}\,-\, \nabla_{\mu}
\nabla^{\nu}R_{\gamma \lambda \varrho \sigma} R^{ \gamma \lambda \varrho \sigma} \right.\right.
\nonumber \\
&&\left.\left.\,+\, 2 R \nabla^{\gamma} \nabla^{\lambda}R_{\gamma \ \lambda \mu}^{\ \nu} \,+\, 2 \Box R_{\gamma \lambda \varrho \sigma} R^{\gamma \lambda \varrho \sigma}\delta_{\mu}^{\ \nu}\,- \
{1\over 2} R R_{\gamma \lambda \varrho \sigma} R^{\gamma \lambda \varrho \sigma} \delta_{\mu}^{\ \nu}\,+\, R_{\mu}^{\ \ \nu} R_{\gamma \lambda \varrho \sigma} R^{\gamma \lambda \varrho \sigma}\,+\,
2 R R_{\gamma \lambda \varrho}^{\ \ \ \nu} R^{\gamma \lambda \varrho}_{\ \ \ \mu}\right.\right.
\nonumber \\
&&\left.\left.\,+\, 2 \nabla_{\lambda} \nabla_{\gamma}R R_{\ \mu}^{\gamma \ \ \lambda \nu}\,+\,2 \nabla_{\lambda} \nabla_{\gamma}R R^{\gamma \nu \lambda}_{\ \ \ \ \mu}\,-\, 2
\nabla_{\varrho}R_{\gamma \lambda}\nabla^{\varrho} R^{\gamma \lambda} \delta_{\mu}^{\ \nu}\,-\,R \nabla^{\gamma}\nabla^{\lambda}R_{\gamma \lambda}\delta_{\mu}^{\ \nu}\right)\,+\,{1\over
2}\eta^{2}(\nabla_{\mu}R\nabla^{\nu}R\right.
\nonumber \\
&&\left.\,+\,\nabla^{\nu}\nabla_{\mu}(\Box R)\,+\,\nabla_{\mu}\nabla^{\nu}(\Box R)\,-\ {1\over 2} \delta_{\mu}^{\ \ \nu}\nabla_{\gamma}R\nabla^{\gamma}R \,-\,2 \Box^{2} R \delta_{\mu}^{\ \
\nu}\,-\,2 \Box R \nabla^{\nu}\nabla_{\mu}R)\,-6\,\eta^{3}(\nabla_{\mu}R\nabla^{\nu}R \right.
\nonumber \\
&&\left.
 \,+\, R \nabla^{\nu}\nabla_{\mu}R\,+\, {1\over 12} R^{3} \delta_{\mu}^{\ \nu}\,-\,R \Box R \delta_{\mu}^{\ \nu}\,-\,\frac{1}{2} R^{2} R_{\mu}^{\ \
 \nu}\,-\,\nabla_{\gamma}R\nabla^{\gamma}R \delta_{\mu}^{\ \nu})\right], \nonumber \\ \label{emTensor2}
\end{eqnarray}
\end{widetext}
where we use \(\Theta=30 - 252 \xi\) and \(\eta=\xi-\frac{1}{6}\). We should stress that a similar calculation was first performed for a general space-time by Matyjasek in reference
\cite{matyjasek}. Our results are somewhat different from that of Matyjasek, but in view of the existence of many tensor identities relating the metric tensor, the Riemann tensor, its contractions
and its covariant derivatives, we expect the two results will be equivalent. The formulas obtained in reference \cite{matyjasek} and those obtained by us, when applied to Schwarzschild,
Reissner-Nordstrom and Kerr space-time shows identical results for the components of the renormalized stress tensor.

In the space-time of a static Black String metric given by (\ref{ds2simp}) simple results were obtained for the renormalized Stress Tensor of massive scalar field showing an arbitrary coupling to
the background gravitational field. After a direct calculation, for the conformal part of the stress tensor we evaluated  in this work the result:
\begin{equation}\label{}
	C_{t}^{\ t}=\frac{11\alpha^{9}\rho^{9}
	-201\alpha ^{3}M^{2}\rho^{3}+1252M^{3}}{2520m^{2}\pi^{2}\alpha^{3}\rho^{9}},
\end{equation}
\begin{equation}\label{}
	C_{z}^{\ z}=C_{\varphi}^{\ \varphi}
	=\frac{11\alpha^{9}\rho^{9}-183\alpha
	^{3}M^{2}\rho^{3}+1468M^{3}}{2520m^{2}\pi^{2}\alpha^{3}\rho^{9}}
\end{equation}

\begin{equation}\label{}
	C_{\rho}^{\ \rho}=\frac{11\alpha^{9}\rho^{9}+189\alpha
	^{3}M^{2}\rho^{3}-308M^{3}}{2520m^{2}\pi^{2}\alpha^{3}\rho^{9}}
	.
\end{equation}
The above components  of the Stress Tensor do not depend in any way  of the coupling constant \(\xi\) because of the constant value of the Ricci scalar in this space-time. The results for the
components of the $D_{\mu}^{\ \ \nu}$ tensor are
\begin{equation}\label{}
	D_{t}^{\ t}=\eta\left[\frac{\alpha^{9}\rho^{9}+112\alpha^{3}\rho
	^{3}M^{2}-704M^{3}}{80\alpha^{3}\rho^{9}\pi^{2}m^{2}}\right]-\frac{9\alpha^{6}\eta^{3}}{2\pi^{2}m^{2}},
\end{equation}
\begin{equation}\label{}
	D_{z}^{\ z}=\eta\left[\frac{\alpha^{9}\rho^{9}+112\alpha^{3}\rho
	^{3}M^{2}-896M^{3}}{80\alpha^{3}\rho^{9}\pi^{2}m^{2}}\right]-\frac{9\alpha^{6}\eta^{3}}{2\pi^{2}m^{2}},
\end{equation}
\begin{equation}\label{}
	D_{\rho}^{\ \rho}=\eta\left[\frac{\alpha^{9}\rho^{9}-112\alpha^{3}\rho
	^{3}M^{2}+192M^{3}}{80\alpha^{3}\rho^{9}\pi^{2}m^{2}}\right]-\frac{9\alpha^{6}\eta^{3}}{2\pi^{2}m^{2}}.
\end{equation}
For the component \(D_{\varphi}^{\ \varphi}\) we have the same value that \(D_{z}^{\ z}\). It is interesting to evaluate the above components of the stress tensor at the event horizon of the black
string. After straightforward calculations we obtain the results:
\begin{equation}\label{}
	T_{t}^{\ t}(\rho_{H})=-\frac{3}{2}\frac{\alpha^{6}\eta}{\pi^{2}m^{2}}\left(\frac{1}{40}+3\eta^{2}\right)+\frac{\alpha^{6}}{140\pi^{2}m^{2}}
\end{equation}
and:
\begin{equation}\label{}
	T_{z}^{\ z}(\rho_{H})=-\frac{3}{2}\frac{\alpha^{6}\eta}{\pi^{2}m^{2}}\left(\frac{1}{20}+3\eta^{2}\right)+\frac{\alpha^{6}}{112\pi^{2}m^{2}}.
\end{equation}
Also we have the relations \(T_{t}^{\ t}(\rho_{H})=T_{\rho}^{\ \rho}(\rho_{H})\) and  \(T_{z}^{\ z}(\rho_{H})=T_{\varphi}^{\ \varphi}(\rho_{H})\). In the general case, all the components of the
renormalized stress energy tensor of the quantized scalar field will be positive at the horizon for the values of the coupling constant satisfying the relation \(3\eta^{3}+\frac{1}{40}\eta <
\frac{1}{210}\). There are some particular cases in which the above relation is always satisfied. The simplest case of the conformal coupling \(\xi=\frac{1}{6}\) and the minimal one are two
important examples. Also for the case \(\xi<\frac{1}{6}\) the components of the quantized scalar field at the horizon are always positive quantities. If we define the energy density as
\(\varepsilon=-T_{t}^{\ t}\), then we can conclude that for the particular cases mentioned above the weak energy condition is violated. However, violations of the weak energy condition for quantum
matter are common, and they are in fact required for a self consistent picture of the Hawking evaporation effect.

The results of this work are expected to be employed to investigate the back-reaction of the quantum scalar field, on the Black String metric. For this purpose, the Einstein  equations for the
metric should be solved after including in them the calculated stress tensor for the Black String solution. Our results of the implementation of this programm will be published in the future.

One of the authors (O.P.F.P) greatly acknowledges M. Chac\'{o}n Toledo and E. R. Bezerra de Mello for helpful discussions. This work was suported by the Abdus Salam International Centre for
Theoretical Physics, Trieste, Italy.

\end{document}